\documentclass[letterpaper,preprint,showpacs]{revtex4}
\usepackage{graphicx}
\begin{document}
\date{\today}
\title{Theory of selective excitation in \\ Stimulated Raman Scattering
}
\author{~S.~A.~Malinovskaya, ~P.~H.~Bucksbaum, and ~P.~R.~Berman}
\affiliation{University of Michigan, FOCUS Center and Department
of Physics,\\ Ann Arbor, MI 48109 }
\begin{abstract}
A semiclassical model is used to investigate the possibility of
selectively exciting one of two closely spaced, uncoupled Raman
transitions. The duration of the intense pump pulse that creates
the Raman coherence is shorter than the vibrational period
 of a molecule (impulsive regime of interaction).
Pulse shapes are
found that provide either enhancement or suppression of particular
vibrational excitations.
\end{abstract}
\pacs{42.65.Dr 42.50.Hz 33.20.Fb 82.53.Kp} \maketitle
\section{Introduction}
Studies of molecular dynamics accompanying excitation,
ionization, or dissociation give rise to questions concerning the
control of specific motion in molecules. Control using ultrafast
lasers and adaptive learning algorithms, e.g., genetic-type
algorithms \cite{We99,Pe01}, allows one to optimize feedback and
achieve predetermined goals. Series of successful experiments
implementing this technique offer proof that it is possible to
exercise quantum coherent control over a range of processes. Two
such experiments employ stimulated Raman scattering in liquid
methanol \cite{Pe01} and gas phase carbon dioxide \cite{We01}. In
the former case a liquid medium is excited with a 100 fs pulse,
which is equivalent to non-impulsive Raman excitation and,
therefore, makes propagation effects important. In the latter case
 the selective Raman excitation of vibrational modes
is achieved with a pulse of duration within a typical molecular
vibrational period. In this case the seed for the Stokes
transitions is within the pump pulse. In the experiment with
methanol the problem of
 intra-molecular couplings of vibrational modes is discussed
using a comparative analysis with experiments on benzene and
deuterated benzene molecules. The controlling optical field may be
constructed using
 coherent quantum control theory, see, e.g. \cite{Sh88,Sh89}.
An alternative approach involves a search for an analytical pulse
function. In \cite{Du02,Or02} an antisymmetric phase function is
proposed, which inverts the sign of the electric field at a given
energy, inducing constructive interference of the off-resonant
components of the spectrum and therefore maximizing the
transition amplitude. Selectivity of the two Raman levels of
pyridine, both of which lie within the bandwidth of the
excitation pulses is achieved in CARS signals by positioning the
phase step at each of the two peaks \cite{Or202}. In \cite{Sc02}
the use of adaptive techniques in FAST CARS is explored for an
identification of related biological objects such as bacterial
spores.

The Raman excitation of a single vibrational mode having frequency
$\omega$ normally requires a pulse with $ \omega \tau \leq 1$,
where $\tau$ is the pulse duration. Optimal excitation may occur
if the intensity is modulated at a frequency that corresponds to
the transition frequency. For any two vibrational modes
$\omega_1$,  $\omega_2$ within the pulse bandwidth, $(\omega_1 -
\omega_2) \tau \ll 1$, the spectral width is too large to resolve
the lines. In the present paper we propose a method for achieving
selective excitation using broadband pulses. In the frequency
domain we
 introduce an intensity envelope that vanishes for
a modulation frequency of the pulse equal to the frequency of
vibration that we would like to suppress. The Fourier transformed
field, when applied to a molecular system, provides negligible
excitation of that particular frequency together with significant
excitation of another frequency. This picture is valid for weak
 fields.
In strong fields the effect
 is not obvious and numerical analysis reveals that, for some
intensities and the designed pulse shape, it is still possible to
optimize one transition versus another.

The paper is organized as follows. In the second section the
problem is formulated based on a model of two, two level systems.
An analytical function for the intensity envelope is proposed and
the equations of motion for the state amplitudes are obtained. The
third section contains results of numerical calculations that
determine the field parameters that lead to selective excitation.
The paper is ended with a short summary.

\section{BASIC FORMALISM}
In our model a molecule is described by two, two-level quantum
systems, each
 representing a normal Raman-active vibrational mode.
The molecular medium is represented by an ensemble
 of such systems with no relaxation
processes taken into account. Each two-level system interacts
with an intense off-resonant femtosecond pulse that initiates
stimulated Raman scattering
 via an off-resonant interaction with a virtual state.
 The duration of this pump pulse is shorter than a typical vibrational
period of a molecule. In this case the frequencies of both
two-level systems are within the bandwidth of the pulse and the
Stokes component of the field is supplied by the same pulse. The
coherently excited ensemble is analyzed by a weak probe pulse, not
considered in this work, applied after a time delay shorter than
the coherence time.
The goal of the present work
 is to determine a pulse shape that provides
selectivity for the excitation of one of the two-level systems.

A semiclassical model of laser-molecule interactions is used. The
model is represented schematically in Fig.1 where $\omega_{21}$ is
the frequency of the first two-level system and $\omega_{43}$
that of the second system. Initially only lower levels $|1>$ and
$|3>$ of both two-level systems are populated, and the
populations of these levels are equal. The time evolution of two,
two-level systems is described in terms of probability
amplitudes, which are written in the interaction representation
\begin{equation}
\dot{a}_j =
i\frac{\Omega_{j}}{4\Delta}\sum_{j'=1}^{4} \Omega_{j'}^{*}
e^{-(\alpha_j-\alpha_{j'}) \omega t} a_{j'}, \qquad
\Omega_{j}=-\frac{\mu_{jb}E_{p0}(t)}{\hbar}.
\end{equation}
Here $\alpha_j \omega$ is the frequency of a single level, such
that, e.g., $(\alpha_2 - \alpha_1) \omega = \omega_{21}$,
$\Omega_j$ is a Rabi frequency, $\mu_{jb}$ is a dipole moment
matrix element, $E_{p0} (t)$ is the pulse envelop, $\Delta$ is the
detuning of the frequency of the pulse from the frequency of the
virtual state $|b>$. Note, that the pulse envelope $E_{p0}(t)$ is
the same for all transitions. The Rabi frequencies may differ
owing to different dipole moment matrix elements.

The system of coupled differential equations (1) is
 derived from the time-dependent
Schr\"odinger equation with Hamiltonian:
\begin{equation}
H =\frac{\hbar}{2} \left( \begin{array}{ccccc} \alpha_1\omega & 0 & 0 & 0
& 2\Omega_{1}cos(\omega_p t)\\
0 & \alpha_2\omega & 0 & 0  & 2\Omega_{2}cos(\omega_p t) \\
0 & 0 & \alpha_3\omega & 0  & 2\Omega_{3}cos(\omega_p t)\\
0 & 0 & 0 & \alpha_4\omega  & 2\Omega_{4}cos(\omega_p t)\\
2\Omega_{1}^*cos(\omega_p t) & 2\Omega_{2}^*cos(\omega_p t) & 2\Omega_{3}^*cos(\omega_p t) & 2\Omega_{4}^*cos(\omega_p t) & E_b\\
\end{array} \right) ,
\end{equation}
where $\omega_p$ is the laser field carrier frequency.
By adiabatically eliminating state $|b>$
 within the rotating wave approximation,
we arrive at Eqs (1).
In this work we discuss the case of uncoupled two-level systems such
 that the probability for the population flow from one system to another via
the external field is zero.
Then Eqs.(1) are represented by two independent systems of coupled differential
equations with two variables.

Coherent excitation of a molecular medium induces a vibrational
coherence $|\rho_{12}|$ or $|\rho_{34}|$. When a probe field is
applied on the 1-b transition, the coherence $|\rho_{12}|$ serves
as a source for the generation of a field on the 2-b transition.
Thus it is of prime interest to calculate $|\rho_{12}|$ and
$|\rho_{34}|$. The goal of this paper is to choose a pump pulse
such that $|\rho_{12}|$ is suppressed while $|\rho_{34}|$ is enhanced.

We propose an analytical function for the intensity envelope which
is included in the dynamical equations (1) for the probability
amplitudes. It is easiest to choose this function in the
frequency domain. To suppress excitation at frequency
 $\omega_{21}$
and enhance excitation at frequency $\omega_{43}$, we choose

\begin{equation}
\tilde{I}(\omega)=I_0 e^{-(\omega-\omega_{43})^2 T^2} \left( 1 -
e^{-(\omega-\omega_{21})^2 {T_1}^2}\right),
\end{equation}
where $T$ and $T_1$ are free parameters.
When the modulation frequency of the pulse
 $\omega$ is equal to $\omega_{43}$ the intensity
approaches its maximum $I_0$ for a sufficiently large parameter
$T_1$,

\begin{equation}
\tilde{I}(\omega_{43}) = I_0 (1 - e^{- \Delta \omega^2 {T_1}^2}),
\qquad \Delta \omega = \omega_{43} - \omega_{21}.
\end{equation}
For $\omega$ equal to $\omega_{21}$ the intensity is zero,
$\tilde{I}(\omega_{21})=0$. The intensity envelope as a function
of frequency is drawn in Fig.2. The frequencies of the two-level
systems are
 $\omega_{21}$=1
and $\omega_{43}$=1.1 in frequency units of $\omega_{21}$.
The intensity of the field at $\omega_{43}$ and frequency region
near $\omega_{21}$ over which $\tilde{I}(\omega)$ is approximately zero
depend on $T_1$. The larger $T_1$, the greater is the selectivity
for suppressing the $\omega_{21}$ frequency.

The inverse Fourier transform of the spectral density (3) is a
complex function. To arrive at a physically acceptable temporal
pulse function, we take the real part of the inverse Fourier
transform, given by

\begin{equation}
I(t)=I_0 \left( (\sqrt2 T)^{-1} e^{-\frac{t^2}{4T^2}}
cos(\omega_{43}t) - (\sqrt2 \tau)^{-1} e^{-\Delta \omega^2 T^2
(1-\frac{T^2}{\tau^2}) - \frac{t^2}{4\tau^2}}
cos((\omega_{21}-\Delta\omega\frac{T^2}{\tau^2}) t) \right),
\end{equation}
where $\tau^2=T^2+T_1^2$.
The expression for the field (5) is inserted in Eqs.(1) for the
calculation of the probability amplitudes.

The solution of Eqs.(1) in the limit of a weak field using
perturbation theory is

\begin{eqnarray}
& a_4= i \frac{\mu_{4b}\mu_{3b}^*}{4\Delta \hbar^2}
\int_{-\infty}^{\infty} I(t) e^{i \omega_{43} t } d t = i
\frac{\mu_{4b}\mu_{3b}^*}{4\Delta \hbar^2} \tilde{I}(\omega_{43})
= i \frac{\mu_{4b} \mu_{3b}^*}{4\Delta \hbar^2} I_0 (1 -
e^{-{\Delta
\omega^2 T_1^2}} ) , \nonumber \\
&\\
& a_2= i \frac{\mu_{2b}\mu_{1b}^*}{4\Delta \hbar^2}
\int_{-\infty}^{\infty} I(t) e^{i \omega_{21} t } d t \sim
e^{-T^2 (\omega_{43} + \omega_{21} )^2} + e^{-\Delta\omega^2 T^2}
e^{-\tau
^2  (2\omega_{21})^2} \approx
 0. \nonumber
\end{eqnarray}
The Fourier transform represented in Eq.(5) is not identical to
Eq.(3), since we took the real part of the Fourier transform to
arrive at (5). It now contains "counter-rotating" terms, which
are small for the chosen pulse shape. Thus, by construction, we
have totally suppressed the 1-2 transition in the weak field
limit. On the other hand, the excitation of the 3-4 transition is
still weak owing to the perturbative nature of the solution.

Polyatomic molecules often possess several or many Raman active
modes with frequencies close enough to be within the bandwidth of the pulse.
In order to enhance a single vibration and suppress other vibrations
the function for
the pulse may be constructed as a product of several terms

\begin{equation}
\tilde{I}(\omega)=I_0 e^{-(\omega-\omega_{43})^2 T^2}\Pi_j \left( 1 -
e^{-(\omega-\omega_{j})^2 {T_1}^2}\right).
\end{equation}

\section{Numerical results}

In this section we discuss
 the results of numerical calculations
based on the exact solution of Eqs.(1). The numerical studies
reveal the influence of the field parameters on the efficiency of
the excitation  of the two-level systems. Parameters for the
system are taken from the experimental data on impulsive
excitation of vibrational modes in the molecular gas $CO_2$
 \cite{Pe01}. In $CO_2$ the frequencies
of two selectively excited Raman modes are 36.8 and 42 THz. The
FWHM of the applied intense pulse is taken equal to 18 THz. In our
calculations the frequency $\omega_{21}$ is set equal to unity;
in these units the frequency $\omega_{43}$ is equal to 1.1. From
experimental data, we estimate that
 the parameter T is about equal
to 3, in frequency units of $\omega_{21}^{-1}$. The intensity of
the field is determined by the parameter $I_0$. The parameter
$T_1$ is related to the width of the spectral dip in
$\tilde{I}(\omega)$ centered at frequency $\omega_{21}$.
Although a value of $T_1 \gg T$ would provide the best
selectivity, the choice for the parameter $T_1$ is strongly
restricted by the requirement that the duration of the applied pulse be
within a typical molecular vibrational period. It turns out that
even for such values of $T_1$, it is possible
 to selectively excite one transition.

We calculated the population distribution and the coherence of the
excited and suppressed two-level systems
 as a
function of $T_1$ as shown in Fig.3. Bold solid and dotted lines
depict the absolute value of the coherence $|\rho_{34}|$ of the
excited system and $|\rho_{12}|$ of the suppressed system.
Populations of the upper levels of both systems are shown by bold
dashed and dot-dashed lines. The intensity of the field
 $I_0$ is $\pi /8$ which corresponds to a weak field in our calculations
 (but not to the perturbative regime).
For the value $T_1 = 13$ the population of levels of the 3-4 system
is 0.25, and the coherence is optimal for the given intensity of
the field and dipole moments, ($\mu_i=1$).
 The duration of the
laser pulse corresponding to this value of the parameter $T_1$ is
about 200 fs which does not satisfy the necessary requirements on
pulse duration. According to Fig.3 for smaller values of the
parameter $T_1$, corresponding to shorter pulses, the coherence
of the 3-4 system is significantly reduced with a simultaneous
increase of the coherence of the 1-2 system. Optimal values for
the coherence of both the 1-2 and 3-4 systems were found for $T_1
\leq T$ through a search over different intensities of the field.

In Fig.4 the coherence is plotted as a function of the intensity
of the field for parameters T=3 and $T_1=3$. Coherence
$|\rho_{34}|$ of the 3-4 system is represented by a bold solid
line and coherence $|\rho_{12}|$ of the 1-2 system by a bold
dashed line; thin lines show
 populations of the upper levels of both
two-level systems. For the intense fields, coherence of the
excited and suppressed systems possess somewhat chaotic
structure. Several values of the intensity, e.g.  $I_0=\pi$ and
$I_0=1.75 \pi$ give rather low coherence of the 3-4 system but
maximum coherence of the 1-2 system.
 A desired solution for maximum coherence of the
excited 3-4 system is achieved for the intensity coefficient $I_0
= 2.08 \pi$.  This is the result of redistribution of population
within that two-level system: half of population is transferred
to the upper level providing maximum coherence. The corresponding
coherence of the 1-2 system at this intensity is nearly zero,
where most population remains in the lower level. For the
intensity $I_0 = 1.4 \pi$ the picture is similar, however the low
coherence of the 1-2 system is due to nearly complete population
transfer to the upper level.

The goal of control of the coherence of two uncoupled two-level
systems is achieved with a pulse shape possessing a broad
spectral dip at the suppressed frequency and a suitably chosen
intensity of the field. This technique allows one to use pulses
of duration $T$ to selectively excite transitions having
frequency separations $\Delta \omega < 1/T$. Our results should
not be taken to imply that one can spectroscopically determine
frequencies to better than the inverse temporal width of the
pulse. On the other hand, if the frequencies are known from
previous measurements, it is possible to suppress one transition
and enhance the other by the method outlined above.

Had we taken a frequency profile centered at $\omega_{43}$ with $T=3$,
the curves
for the coherence and populations of the excited and suppressed
systems would differ qualitatively from those show in Fig.4. The
desired selectivity could not be achieved.

 The time-dependence of the coherence,
populations and the field is shown in Fig.5
 for $T=3, T_1=3 $,
$I_0=1.4 \pi$. The pulse duration is about 50 fs (thin solid
line). It induces oscillations in the population distribution
(thin lines) which lead to oscillations of the coherence of the
two-level systems (bold lines). At long times the coherence and
populations of levels achieve stationary values.

We have carried out some preliminary calculations, including the
coupling between vibrational modes via an external field used as a
control mechanism.
If the direct excitation of a particular vibrational mode is weak
owing to a weak oscillator strength, the excitation may be
enhanced through the coupling to the other Raman active
vibrational modes. Details will be published in \cite{Ma04}.

\section{Summary}

We have presented a semi-classical description of stimulated Raman
scattering involving the selective excitation of one of two
closely spaced vibrational modes. The dynamics is described in
terms of probability amplitudes and depends on the shape of a
femto-second laser pulse where duration is shorter than a typical
molecular vibrational period. We propose an analytical function for the
shape of the intensity envelop of the pulse that allows for the
selective excitation of a predetermined vibrational motion with
simultaneous suppression of an unfavorable one. The pulse leads to
maximum coherence of a desired vibrational transition and
consequently to maximum gain into Raman side bands when a probe
pulse is applied. This pulse function may be used as an initial
guess for the control of bond excitation in chemistry as well as
a fit function within an adaptive learning algorithm.

\section*{ACKNOWLEDGMENTS}

The authors acknowledge financial support from the National
Science Foundation (No. PHY-9987916) through the Center for
Frontiers in Optical Coherent and Ultrafast Science (FOCUS).

{}

\newpage
\begin{figure}
\includegraphics[width=10cm]{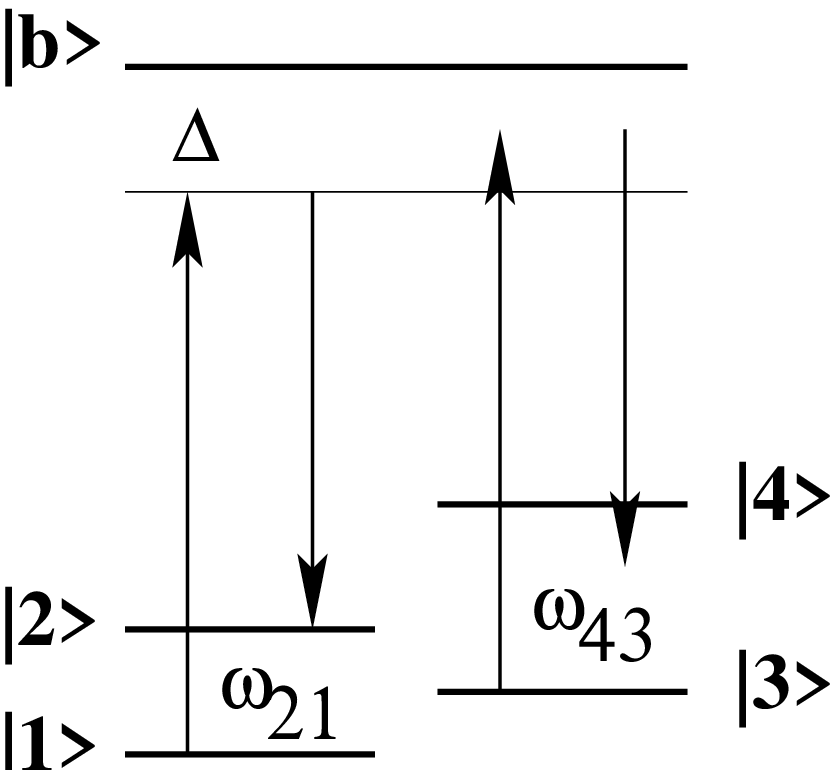}
\caption{Schematic picture of a model system consisting of two,
two-level systems having frequencies $\omega_{21}$ and
$\omega_{43}$. Initially, the lower levels are populated evenly.
The uncoupled transitions are driven by an off-resonant
femtosecond pulse.\\}
\end{figure}

\newpage
\begin{figure}
\includegraphics[width=10cm]{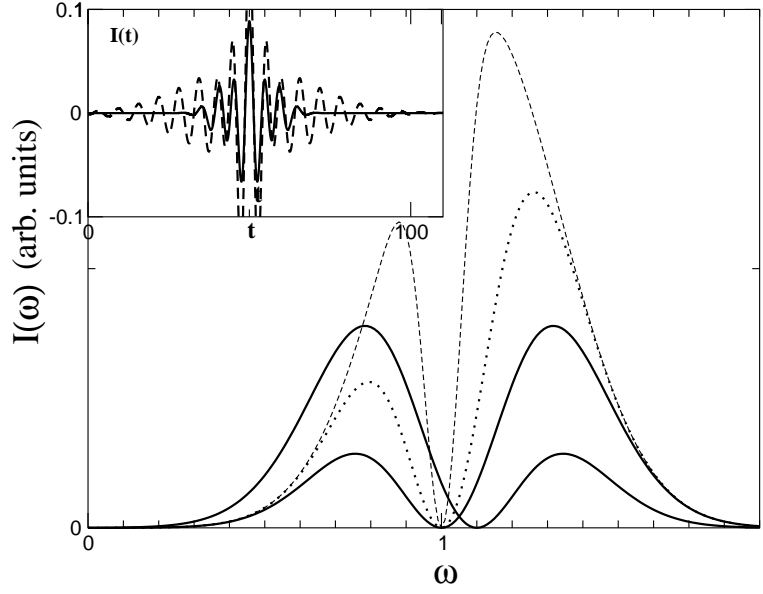}
\caption{ Intensity spectral profile as a function of frequency
for $T_1 = 10,5,3 $
 (dashed, dotted, and solid lines).
In the insert the intensity envelop as a function of time is
presented for $T_1 = 10, 3$. All frequencies are in units of
$\omega_{21}$ and times in units of $\omega_{21}^{-1}$.\\}
\end{figure}

\newpage
\begin{figure}
\includegraphics[width=10cm]{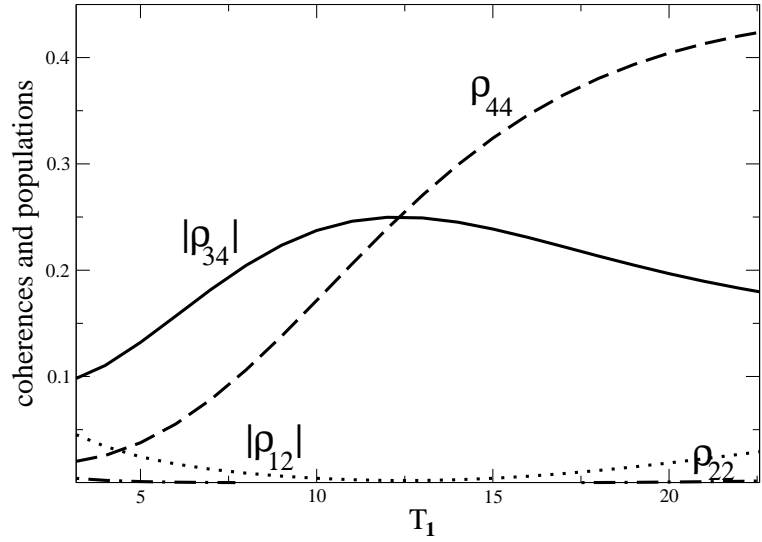}
\caption{ Coherence and populations of upper levels of the 3-4 and
1-2  two-level systems as a function of $T_1$ for $\omega_{21}=1,
\omega_{43}=1.1 $, and $ I_0=\pi/8$.\\}
\end{figure}

\newpage
\begin{figure}
\includegraphics[width=10cm]{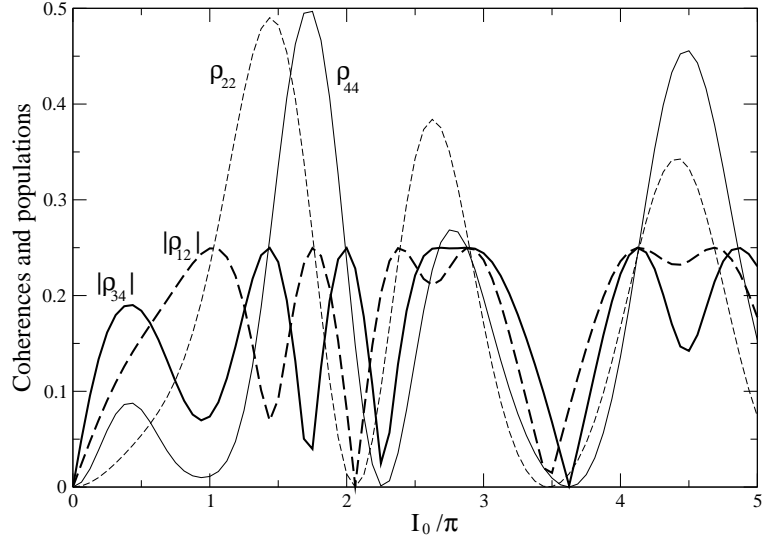}
\caption{ Intensity dependence of the coherences $|\rho_{12}|$ and
$|\rho_{34}|$ and upper states populations $\rho_{22}$ and
$\rho_{44}$. Maximum coherence of the 3-4 system and negligibly
small coherence of the 1-2 system are observed for $I_0=1.4$ and
$2.08$ in the intensity region shown with
 $T=3 ,T_1=3.$\\}
\end{figure}

\newpage
\begin{figure}
\includegraphics[width=10cm]{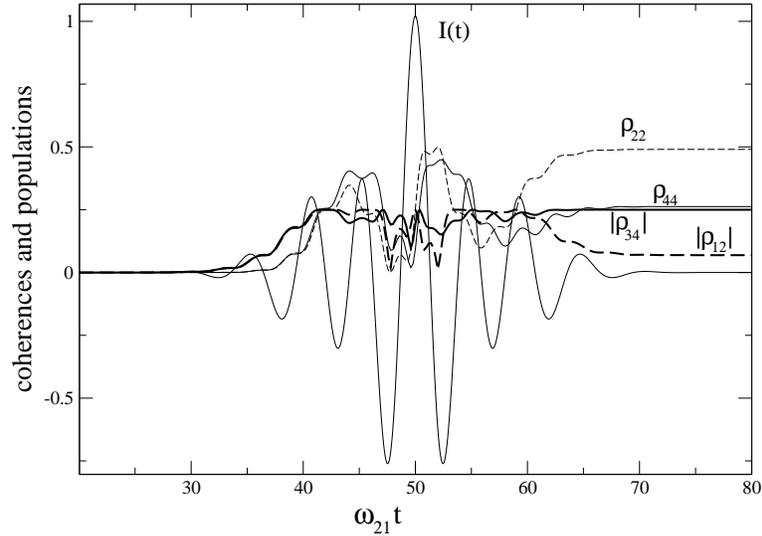}
\caption{ Time evolution of $|\rho_{12}|$, $|\rho_{34}|$,
$\rho_{22}$, and $\rho_{44}$ for T=3, $T_1=3 $, and $ I_0=1.4
\pi$. The pulse intensity I(t), in arbitrary units, is also shown
in the figure.\\}
\end{figure}
\enddocument